\documentclass[
preprint,
 amssymb, amsmath,
 aps,cha,
titlepage,
longbibliography
]{revtex4-1}

\usepackage{graphicx}
\usepackage{epstopdf}
\usepackage{mathtools}
\usepackage[percent]{overpic}
\usepackage[colorlinks=true,linkcolor=blue]{hyperref}
\usepackage{soul}

\usepackage{subcaption}
\DeclareCaptionFormat{cont}{#1 (cont.)#2#3\par}

\draft

\begin{document}

\title{Rebounds of deformed cavitation bubbles}

\author{Outi Supponen}
\affiliation{Department of Mechanical Engineering, University of Colorado, Boulder, USA}
\author{Danail Obreschkow}
\affiliation{International Centre for Radio Astronomy Research, University of Western Australia, Australia}
\author{Mohamed Farhat}
\affiliation{Laboratory for Hydraulic Machines, Ecole Polytechnique F\'ed\'erale de Lausanne, Switzerland}

\date{\today}

\begin{abstract}
Presented here are experiments clarifying how the deformation of cavitation bubbles affects their rebound.
Rebound bubbles carry the remaining energy of a bubble following its initial collapse, which dissipates energy mainly through shock waves, jets, and heat.
The rebound bubble undergoes its own collapse, generating such violent events anew, which can be even more damaging or effective than at first bubble collapse.
However, modeling rebound bubbles is an ongoing challenge because of the lack of knowledge on the exact factors affecting their formation.
Here we use single-laser-induced cavitation bubbles and deform them by variable gravity or by a neighboring free surface to quantify the effect of bubble deformation on the rebound bubbles.
Within a wide range of deformations, the energy of the rebound bubble follows a logarithmic increase with the bubble's initial dipole deformation, regardless of the origin of this deformation.
\end{abstract}

\maketitle
\section{Introduction}

The collapse of cavitation bubbles has attracted interest in the research community because of its substantial damage potential.
They are harmful typically to hydraulic machines such as turbines~\cite{Escaler2006}, pumps~\cite{Spraker1965}, and ship propellers~\cite{Terwisga2007}, but also beneficial in many biomedical applications, such as lithotripsy using shock waves or high-intensity focused ultrasound~\cite{Coleman1987,Ikeda2006} or sonoporation~\cite{Marmottant2003,Ohl2006,Brennen2015}, in cleaning applications~\cite{Fernandez2012}, and in micropumping \cite{Dijkink2008}, among other applications.
The pursuit of controlling the damaging properties of cavitation and of understanding the underlying physics has motivated numerous studies to look at simplified case studies, especially involving the first collapse of single cavitation bubbles in various conditions.
A great deal is indeed known about these bubbles and how they are able to emit microjets moving at hundreds of meters per second~\cite{Blake1987,Philipp1998}, shock waves with peak pressures reaching thousands of atmospheres~\cite{Vogel1988,Holzfuss1998,Pecha2000}, and luminescence, i.e.,\ light emission due to the extreme heating of the bubble interior reaching thousands of degrees in temperature~\cite{Brenner2002,Baghda2001}.
Our understanding of these processes comes from the combination of experimental studies, often using laser- \cite{Vogel1988,Robinson2001,Sankin2005,Reuter2018}, spark- \cite{Obreschkow2006,Zhang2015,Zhang2016}, or ultrasound-induced~\cite{Matula2000,Brujan2005} cavitation bubbles; numerical studies typically using boundary integral methods~\cite{Taib1983,Robinson2001,Tomita2002,Zhang1993,Pearson2004b,Wang2016} and domain methods~\cite{Johnsen2009,Hsiao2014,Chahine2015,Koukouvinis2016,Koukouvinis2016b}; and analytical studies, most of them considering bubbles collapsing spherically~\cite{Rayleigh1917,Akhatov2001,Hauke2007,Trilling1952,Keller1980,Prosperetti1987,Fuster2010} but some also tackling non-spherical bubble shapes, for example, by using the concept of Kelvin impulse~\cite{Blake1988,Blake2015}.

Non-sphericity adds a lot of complexity to the modeling of the bubble collapse.
We also do not precisely know the composition, the quantity, the source, or the behavior of the gaseous contents of a typical cavitation bubble in its most extreme collapse conditions, making accurate modeling even more challenging.
During the final stages of the collapse, these bubble contents are violently compressed and act as a spring, making the bubble interface bounce off them and forming thus a rebound bubble. 
The rebound's characteristics (for instance, its geometry or distance to a surface) often vary from those of the first bubble oscillation, and its ensuing collapse can lead to effects that are comparable or even more damaging. 
It is therefore important to understand the physics underlying multiple bubble oscillations, which is not yet fully understood.
Many theoretical and numerical studies rely on experimental observations to model the rebound bubble, of which the formation depends on multiple factors, such as gas inside the bubble, mass transfer, phase change, possibly chemical reactions, and, as we further detail in this paper, bubble deformation.
Here, we present an experimental study conducted explicitly on the rebound formation of single cavitation bubbles, aiming to understand the direct role of the bubble's deformation on them.

As has previously been established by numerous studies, with only a fraction listed here, the main microjet characteristics such as the jet speed, impact timing, bubble displacement, jet volume~\cite{Supponen2016,Obreschkow2011}, as well as the shock wave strength (peak pressure and energy)~\cite{Vogel1988,Supponen2017a} and the luminescence energy~\cite{Ohl2002,Supponen2017b}, vary as a function of the bubble's level of deformation. 
We quantify this with an anisotropy parameter $\zeta$, which represents a dimensionless equivalent of the linear momentum of the liquid accumulated at the bubble's collapse (Kelvin impulse)~\cite{Blake1988} and which is defined for different sources of bubble deformation, such as near surfaces and uniform pressure gradients~\cite{Supponen2016,Obreschkow2011}. 
With an increasing bubble deformation, i.e., with increasing $\zeta$, the jet speed decreases and so does the resulting water hammer pressure and shock energy (up to $\zeta\sim0.1$)~\cite{Supponen2017a}. 
It is therefore reasonable to expect that the rebound energy also varies with $\zeta$, which has been implicitly suggested by previous studies~(e.g., Refs.~\cite{Zhang2015,Wang2016}).
Here, we want to quantify such variations.

For bubbles collapsing in a pressure gradient, such as the one induced by gravity, the deformation-quantifying anisotropy parameter reads $\zeta = \nabla p R_{0} \Delta p^{-1}$, where $\nabla p$ is the pressure gradient ($=\rho g$ for gravity, where $\rho$ is the liquid density and $g$ is the gravitational acceleration), $R_{0}$ is the maximum bubble radius, and $\Delta p$ is the driving pressure ($=p_{0}-p_{v}$, where $p_{0}$ is the static pressure of the liquid at the height of the bubble and $p_{v}$ is the vapor pressure).
For gravity-induced deformations, it is similar to the previously introduced buoyancy parameter~\cite{Gibson1968,Blake1988,Zhang2015}.
The the microjet is always directed against the pressure gradient (e.g., upward for gravity).
For bubbles collapsing near a flat free or rigid surface, the anisotropy parameter can be expressed in terms of the dimensionless stand-off distance of the bubble to the surface $\gamma = h/R_{0}$ (where $h$ is the smallest distance between the initial bubble center and the surface) as $\zeta=0.195\gamma^{-2}$~\cite{Supponen2016}.
The derivation of this relation is based on determining an equivalent uniform pressure gradient (first term of the Taylor expansion of any anisotropic pressure field) yielding the same Kelvin impulse as a neighboring boundary \footnote{Note that a simpler way to approximate the relationship between $\zeta$ and $\gamma$ for a bubble near a rigid surface would be to compute the pressure gradient $\nabla p$ in $\zeta$ by combining the unsteady Bernoulli equation and an image source with potential flow $\phi=R^{2}\dot{R}r^{-1}$, where $r$ is the radial distance away from the source. At the maximum bubble radius, when the effect of the boundary is assumed the strongest, the non-dimensionalized pressure gradient produced by the image at the bubble's center yields a similar relation $\zeta=0.25\gamma^{-2}$.}.
The ensuing microjet for a neighboring rigid and free surface is directed towards and away from the boundary, respectively.
In the following, we measure the energy of the rebound bubble as a function of the bubble's initial deformation induced either by gravity or by a neighboring free surface.

\begin{figure}
\begin{center}
\begin{overpic}[width=.55\textwidth]{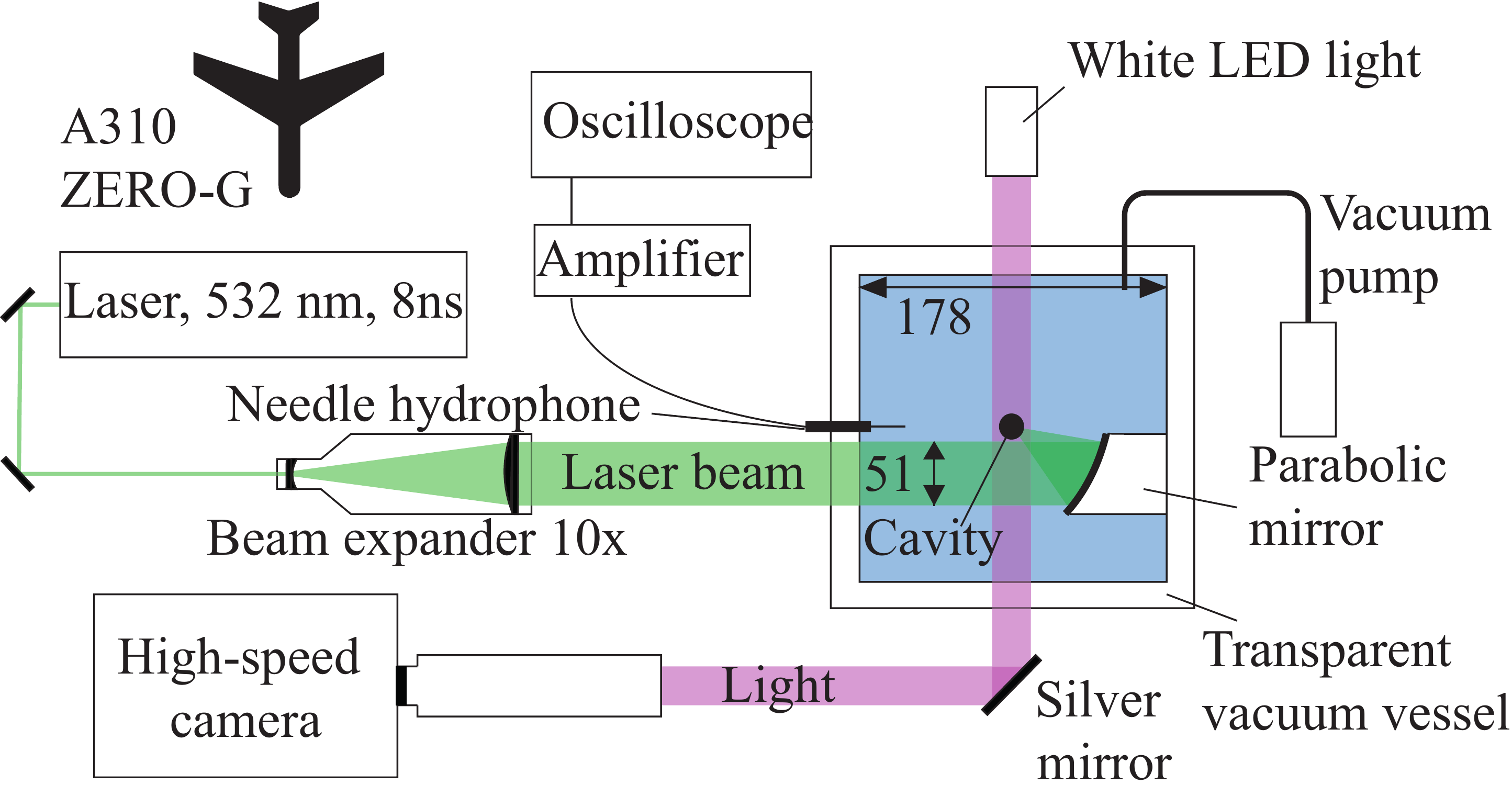}
\put (1,50) {(a)}
\end{overpic}
\caption{Schematic of the experimental setup and direction with respect to aircraft. The dimensions are given in millimeters.} 
\label{fig:fig1}
\end{center}
\end{figure}

\section{Experimental setup}

The central elements of our experimental setup are illustrated in the schematic of Fig.~\ref{fig:fig1}. 
The cavitation bubble is generated by means of a pulsed laser (Nd:YAG, 532~nm, 8~ns), focused by an immersed parabolic mirror in the middle of a cubic test chamber (18$\times$18$\times$18~cm) filled with demineralized and partially degassed water.
The static pressure inside the test chamber is adjusted using a vacuum pump, and the dissolved oxygen concentration is monitored by means of an optical oxygen sensor placed near the bottom of the chamber.
The dissolved oxygen concentration is subsaturated and fairly constant ($6.0\pm0.4$~mg/L) for all measurements made at static pressures above 30~kPa, but reaches saturation and decreases accordingly to Henry's law at lower pressures down to $2.8\pm$0.2~mg/L at 10~kPa.
The initial laser-induced plasma is pointlike and thereby generates initially so spherical bubbles that, when they collapse without any external sources of deformation, we may easily observe luminescence from a single collapse in images and in measured spectra~\cite{Supponen2017b}.

The timing of the bubble generation and the first, second, and third bubble oscillations are recorded by a preamplified piezoelectric needle hydrophone (Precision Acoustics, 75~$\mu$m sensor), which detects the passage of the corresponding shock waves and is connected to an oscilloscope sampling at 100-MHz frequency.
The hydrophone needle is sufficiently thin (0.3~mm diameter) and far away (33~mm from bubble center) that we assume it to have a negligible effect on the bubble dynamics.
We also assume that the test chamber is big enough for the reflections of the shock waves from its boundaries to have a negligible effect on the bubble dynamics, since the shock waves propagate spherically and therefore their peak pressure rapidly decreases.
Additionally, we observe the bubble dynamics through high-speed imaging (Shimadzu HPV-X2, up to 10~million~frames/s), where the instantaneous radius of the bubble is obtained by fitting a circle on the obtained images using automated image processing.

By adjusting the laser energy and the static pressure inside the test chamber, we can create vapor bubbles of maximum radii within the range $R_{0}=0.5$--$10$~mm collapsed by a driving pressure within the range $\Delta p=8$--$100$~kPa.
The initial bubble energy here varies between 0.1 and 28~mJ [where the energy is computed as the potential energy of the bubble as its maximum size, i.e., $E_{\rm 0} = (4\pi/3)R_{\rm 0}^{3}\Delta p$].
The experiments involving bubbles deformed by gravity were conducted during the 67th European Space Agency (ESA) parabolic flights in 2017, where we could vary the perceived gravitational acceleration within the range $0$--$2\times9.81$~ms$^{-2}$.
The advantage of deforming the bubbles by gravity is both that (i)~such a uniform pressure gradient approximates to first order any pressure field and that (ii)~the rebound bubbles feel the same pressure gradient as the first bubble oscillation, which is not the case for bubbles collapsing near surfaces, since they unavoidably migrate toward or away from that surface, for solid or free surfaces, respectively.
Additionally, we deform the bubbles by introducing a free surface near the bubble at a distance in the range $h=2$--$32$~mm in order to reach higher levels of deformation and to uncover the rebound's dependence on the source of deformation.
Further details on the experimental setup may be found in Refs.~\cite{Obreschkow2013,Supponen2017a}.

\section{Results}

\begin{figure}
\begin{center}
\begin{overpic}[width=\textwidth, trim=0cm 1.5cm 0cm 0cm, clip]{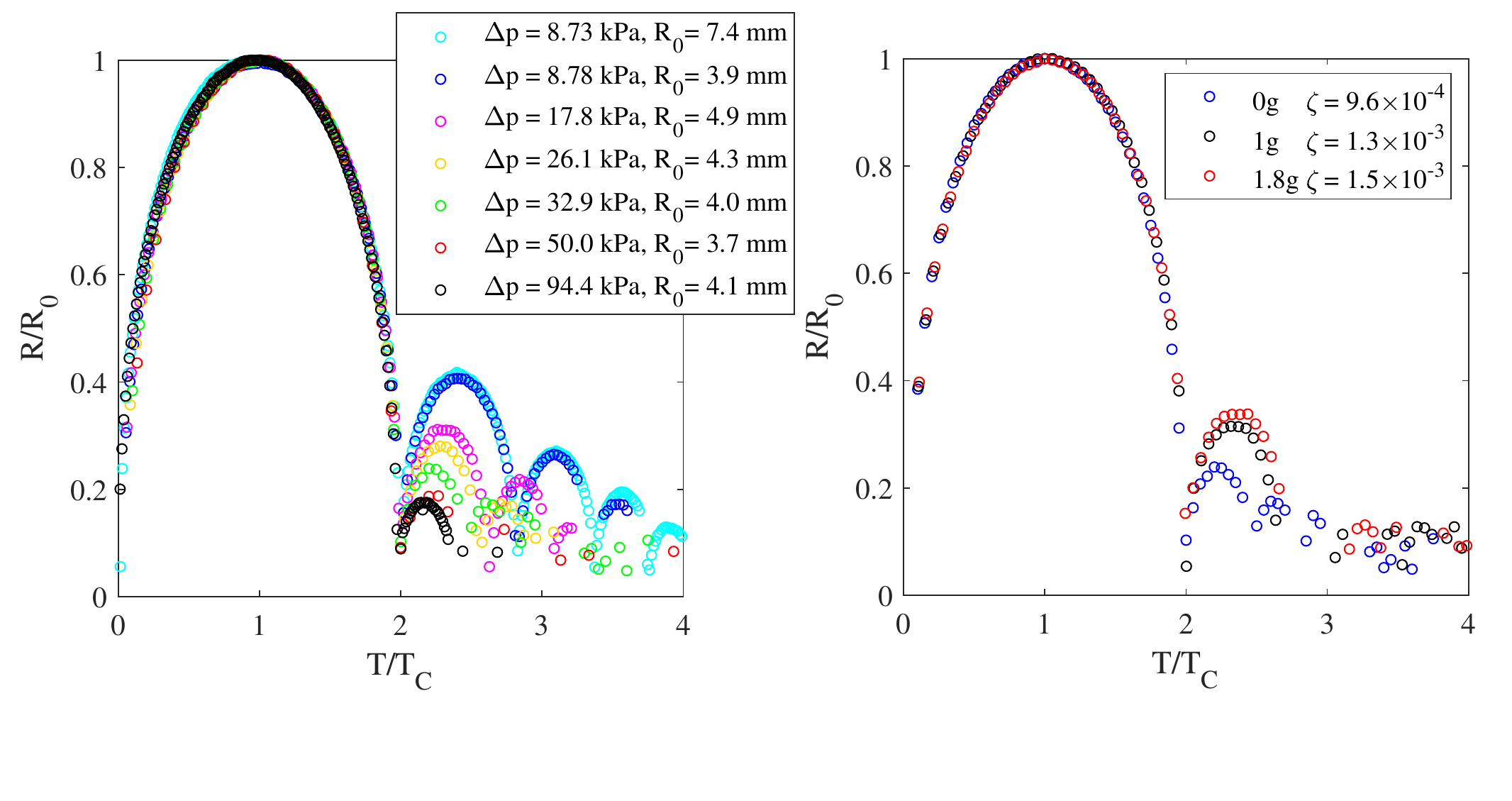}
\put (9,8) {(a)}
\put (61,8) {(b)}
\end{overpic}
\caption{Bubble radius evolution in normalized coordinates [radius $R$ normalized to maximum bubble radius $R_{0}$ and time $T$ normalized to Rayleigh collapse time $T_{C}=0.915R_{0} (\rho/\Delta p)^{1/2}$] as extracted from the high-speed images for bubbles collapsing (a)~spherically in microgravity (0~$g$) with variable driving pressures and (b)~at variable gravity levels: 0, 1, and 1.8~$g$, where $g = 9.81$~ms$^{-2}$, with  $R_{0}\approx 3.9\pm0.2$~mm and at a driving pressure $\Delta p=33$~kPa.} 
\label{fig:fig2}
\end{center}
\end{figure}

Figure~\ref{fig:fig2} shows, in normalized coordinates, the radial evolution of cavitation bubbles at different conditions as extracted from the images. 
The bubbles in Fig.~\ref{fig:fig2}(a) collapse highly spherically in microgravity, and their rebound dynamics show differences depending on the driving pressure $\Delta p$.
At $\Delta p =$ 94~kPa and 50~kPa, the rebound bubble's maximum radius is less than 20\% of the initial bubble's maximum radius.
This means that the rebound has taken less than 1\% of the bubble's initial energy.
However, similarly sized bubbles collapsing at a lower driving pressure $\Delta p$ yield a bigger rebound.
At $\Delta p =$ 9~kPa, the rebound bubble's radius is up to 40\% of the initial bubble's radius, meaning that 7\% of the bubble's initial energy has gone to the rebound.
This finding agrees with the model by Tinguely \emph{et al.}~\cite{Tinguely2012}, where the fraction of a spherically collapsing bubble into the rebound and shock waves was shown to vary with $\Delta p$ (as well as the non-condensible gas, heat capacity ratio, speed of sound, and density of water), but to be independent of the bubble's maximum radius $R_{0}$.
The ratio between the initial bubble radius and the maximum radius of the rebound bubble was also found to be constant when only varying $R_{0}$ in the mathematical model by Akhatov \emph{et al.}~\cite{Akhatov2001}.
Figure~\ref{fig:fig2}(a) confirms such independence of $R_{0}$ (at least in the parameter space covered here); at the same driving pressure $\Delta p = 9$~kPa, two bubbles of very different sizes $R_{0}=7.4$~mm and $R_{0}=3.9$~mm (and thereby also different collapse times) show almost identical behavior over multiple oscillations, as implied by their curves being superimposed up to the third oscillation.

The bubbles in Fig.~\ref{fig:fig2}(b) collapse in otherwise identical conditions but at three different levels of gravity.
There is a clear effect of gravity on the rebound formation, as the bubbles collapsing in normal and hypergravity form a relatively larger rebound compared to the bubble collapsing in microgravity.
A small variation in the bubble's maximum radii $R_{0}$ was observed at the different levels of gravity despite the same laser output energy, with slightly smaller bubbles produced in higher gravity levels due to the additional static pressure at higher $g$.
The nonzero $\zeta$ for the bubble in micro-gravity comes from the nearest surface (parabolic mirror at 55~mm from bubble), which we take into account when calculating $\zeta$.

\begin{figure}
\begin{center}
\begin{overpic}[width=.77\textwidth]{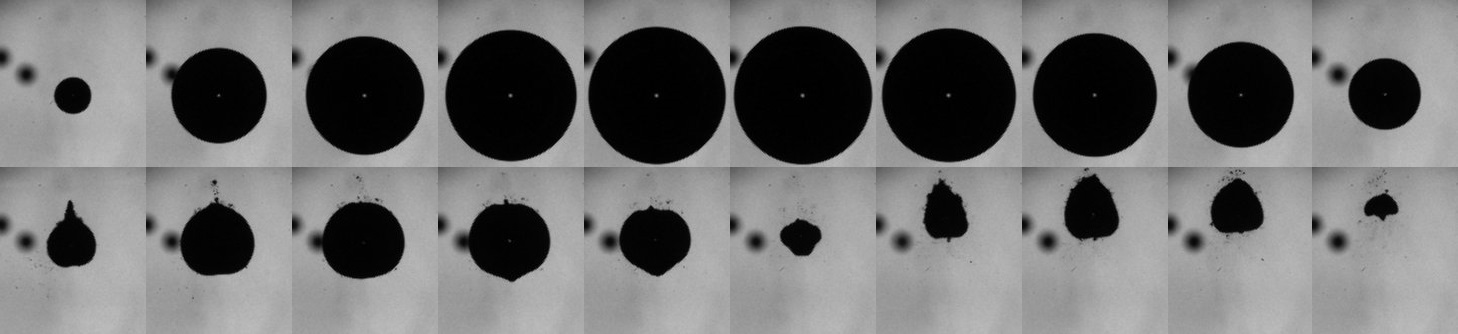}
\put (1,20) {\textcolor{white}{(a)}}
\end{overpic}
\begin{overpic}[width=.77\textwidth, trim=1.3cm 0cm 1.5cm 0cm, clip]{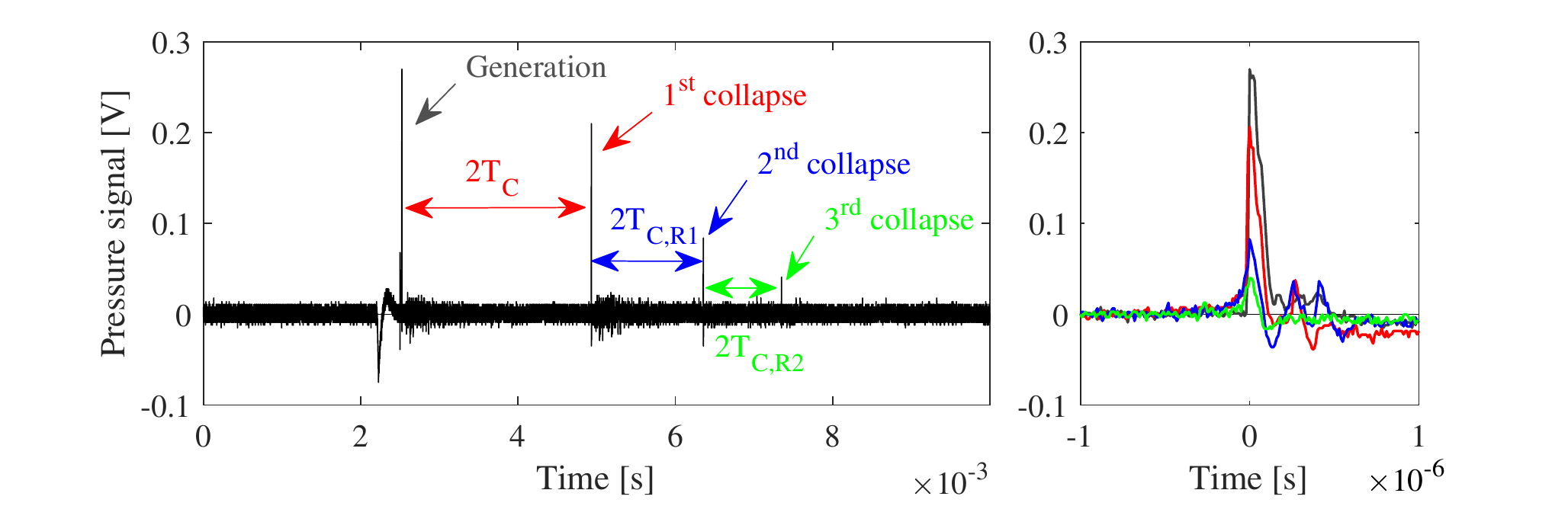}
\put (9,30.5) {(b)}
\put (74,30.5) {(c)}
\end{overpic}
\caption{(a)~Visualization of the first three oscillations of a bubble deformed by the gravity-induced pressure gradient. The frame rate is 250~$\mu$s, exposure time 100~ns, and the maximum bubble radius is 4.4~mm ($\Delta p=11$~kPa, $\zeta=0.0059$, $g = 1.6\times9.81$~ms$^{-2}$). (b)~Full raw pressure signal corresponding to the bubble in panel (a), as measured by the hydrophone and recorded by the oscilloscope. (c)~Enlargement of the shock wave signals, aligned in time by their maximum amplitude to superimpose the peaks.}
\label{fig:fig3}
\end{center}
\end{figure}

Figure~\ref{fig:fig3}(a) shows an image sequence of multiple oscillations of a bubble deformed by gravity.
A microjet, directed upward against the gravity-induced pressure gradient, is visible during the first rebound.
The energy of the rebound bubble is computed as $E_{\rm R} = (4\pi/3)R_{\rm R}^{3}\Delta p$, where $R_{\rm R}$ is the radius of the spherical equivalent of the bubble, i.e., with the same collapse time $T_{\rm C,R} = 0.915 R_{\rm R} (\rho/\Delta p)^{1/2}$.
We compute the rebound energy this way rather than only using the bubble radii obtained from images, since it is difficult, using the latter, to accurately determine the rebound energy of deformed and jetting bubbles with complex geometries.
The collapse time for the first rebound bubble, $T_{\rm C,R1}$, is obtained from the hydrophone signal as the half time between the first and second collapse shock peak pressures, as displayed in Fig.~\ref{fig:fig3}(b).
Correspondingly, the equivalent radius of the second rebound bubble is computed from the measured time difference between the second and the third collapse shock timings.
Each rebound generally emits shocks with decreasing amplitudes, as shown in Fig.~\ref{fig:fig2}(c).
We estimate both the bubble migration toward or away from the hydrophone during each oscillation, and the choice of collapse shock in case of multiple shock wave emission~\cite{Supponen2017a}, to cause a negligible nanosecond-scale error on the measured collapse time.
The second bubble collapse indeed emits multiple shock waves, which is manifested by the numerous peaks in the corresponding signal in Fig.~\ref{fig:fig3}(c).

\begin{figure}
\begin{center}
\includegraphics[width=\textwidth, trim=1cm 10.2cm 1.3cm 10.2cm, clip]{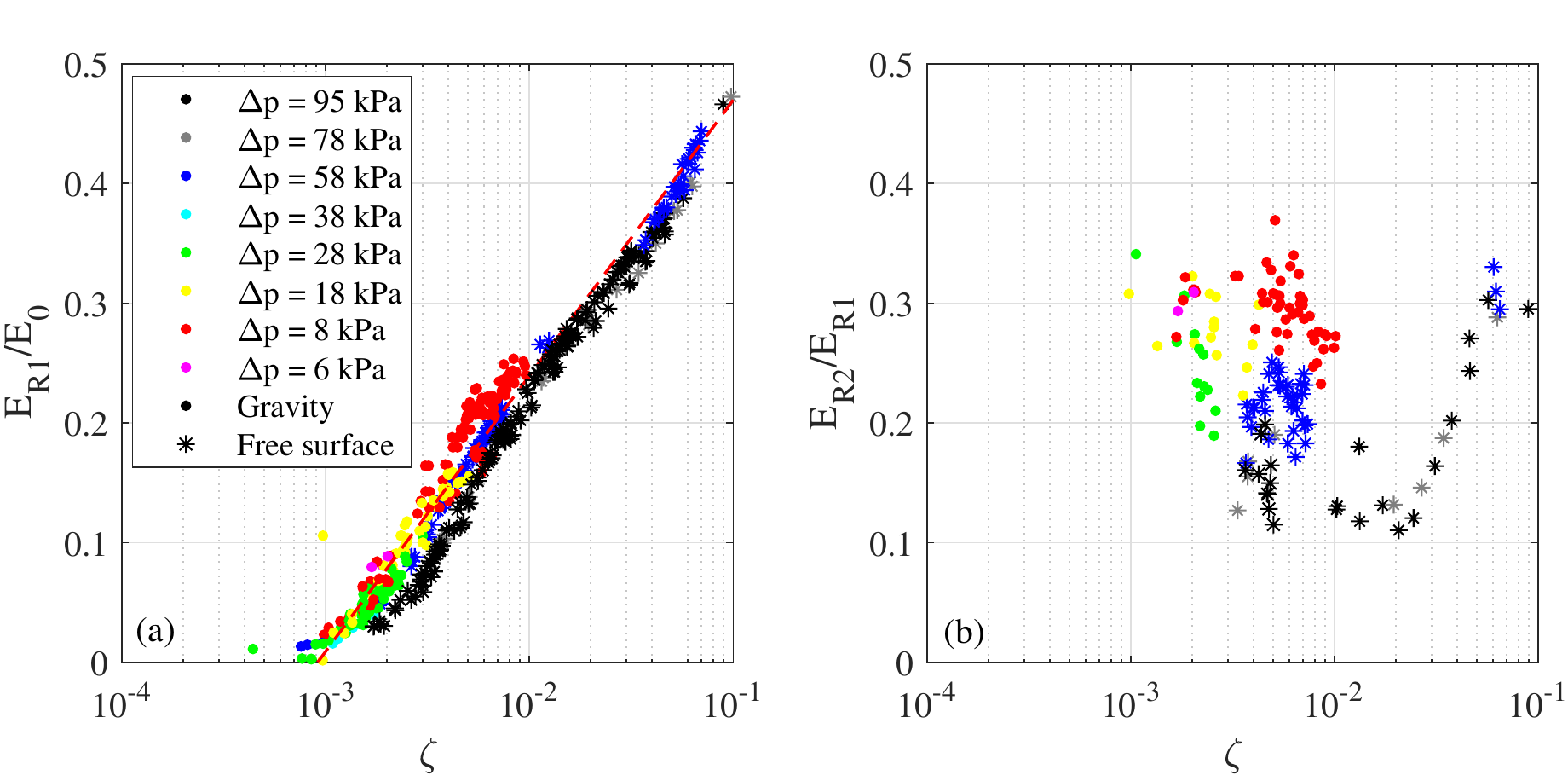}
\caption{(a)~Energy of the first rebound bubble normalized to the initial bubble energy $E_{\rm R1}/E_{0}$, and (b)~energy of second rebound bubble normalized to the first rebound energy $E_{\rm R2}/E_{\rm R1}$, as a function of the anisotropy parameter $\zeta$. The bubbles are deformed by the gravity-induced pressure gradient (with $R_{0}=1.2$--$6.2$~mm, $\Delta p=8$--$83$~kPa, and $g=0.0$--$2.1$~ms$^{-2}$) and by a near free surface (with $R_{0}=1.3$--$4.8$~mm, $\Delta p=50$--$96$~kPa, and $h=2$--$32$~mm). The dashed lines display the logarithmic best fit $E_{\rm R1}/E_{0} \approx 0.1\ln{\zeta}+0.7$. Each data point corresponds to a single collapse event.}
\label{fig:fig4}
\end{center}
\end{figure}

The relative energy of the first and second rebound bubbles as a function of $\zeta$ for bubbles deformed by gravity and by a nearby free surface are shown in Figure~\ref{fig:fig4}.
The energy fractions are relative to the preceding bubble oscillation.
It should also be noted that, for simplicity, $\zeta$ is kept as the \emph{initial} anisotropy parameter determined for the first bubble collapse only.
The actual anisotropy felt by the rebound bubbles is different because of the accumulated momentum during the rebound phases.
The rebound energies measured from the hydrophone signals and the high-speed recordings are generally in good agreement, with their differences here remaining below 5\% at $\zeta<10^{-2}$.
At higher levels of deformation, $\zeta>10^{-2}$, the geometry of the rebound bubble deviates too much from a sphere that fitting a circle to it to extract a radius is no longer accurate, and consequently the rebound bubble size extracted from the high-speed recording is generally underestimated.

The first rebound bubble gains energy with increasing $\zeta$, following a logarithmic increase [Fig.~\ref{fig:fig4}(a)].
For highly spherical collapses in our experiment, most of the bubble energy (80-90\%) is carried away by a shock wave~\cite{Supponen2017a} and less than 10\% goes to the formation of the rebound. 
There are almost no data points for rebound energies computed from the hydrophone signal in Fig.~\ref{fig:fig4} at $\zeta <10^{-3}$, as the peak pressure in the hydrophone signal corresponding to the rebound bubble collapse shock was not detectable for such spherical bubble collapses.
As the bubble's spherical symmetry is broken, the shock wave energy decreases~\cite{Supponen2017a}, the kinetic energy of the nonconverging motion of the liquid increases, and the rebound bubble energy increases.
The shock wave emitted from the rebound bubble's collapse consequently also increases in amplitude with increasing $\zeta$ and becomes detectable by the hydrophone.

Within the range $10^{-3}<\zeta<10^{-1}$, the best fit to the rebound energies of bubbles deformed both by gravity and a free surface is a logarithmic function of the form $E_{R1}/E_{0} = 0.1\ln{\zeta}+0.7$, with a high determination coefficient, $R^{2}=0.97$.
It is interesting that the data points collapse so well onto this single curve despite the large variation in bubble radii, driving pressures, and the two different sources of bubble deformation covered in the experiment.
This suggests that the effect of bubble deformation on the rebound dominates over the effects of driving pressure and dissolved air concentration, which varies in our experiment with $\Delta p$ at low pressures.
It also implies that the rebound energy is practically independent of the source of deformation within this range, despite the collapse shock energies and the geometries at the last stages of the collapse being rather different~\cite{Supponen2016,Supponen2017a}.
Below this range ($\zeta<10^{-3}$, or $\gamma >14$), the bubbles are collapsing in the \textit{weak jet regime}~\cite{Supponen2016}, meaning that no visible jet is produced during the bubble's rebound but the whole collapse appears mostly spherical, and the rebound bubble is relatively small ($E_{R1}/E_{0}<10\%$).
Above this range ($\zeta>10^{-1}$, or $\gamma<1.4$), the bubbles are collapsing in the \textit{strong jet regime}~\cite{Supponen2016}, for which we have no rebound measurements.
This is because bubbles in our experiment cannot reach this regime with gravity-induced deformations, while the dynamics of bubbles deformed by such a near free surface are too strongly affected by the physical presence of the free surface (risk of bubble bursting, shock wave reflections, etc.), making the rebound energy quantification challenging.
However, the rebound energy at such high levels of deformation likely does not follow the logarithmic trend any longer.
It has been observed that for bubbles collapsing near a rigid surface at $\zeta > 0.2$ ($\gamma<1$)~\cite{Vogel1988}, the collapse shock wave energy increases with decreasing distance to the surface, meaning that less energy remains available for the rebound.

The second rebound bubble collapse also becomes detectable on the hydrophone signal only at $\zeta>10^{-3}$.
As shown in Fig.~\ref{fig:fig4}(b), the second rebound takes between 20\% and 35\% of the first rebound's energy for bubbles deformed by gravity and between 10\% and 30\% for bubbles deformed by a free surface.
No clear variation of this ratio as a function of the initial $\zeta$ is found. 
The strong scatter of the data is likely due to the strong dependence of the second collapse on the uncontrollable microvariations in the initial conditions. 
Initial perturbations on the bubble's sphericity become amplified over the bubble oscillations, and eventually damping fission instabilities occur~\cite{Brenner2002}.
However, visible bubble splitting for reasons other than the microjet is not observed in the high-speed recordings generally until the fourth bubble oscillation.
This observation varies with the bubble deformation and size of the first rebound, and it is more troublesome to determine whether the tiny oscillations following a spherical bubble collapse remain as a single entity [note that such spherical bubble collapses are not included in Fig.~\ref{fig:fig4}(b)].
There are differences between the second rebounds of bubbles deformed by gravity and free surface, with generally lower energy fractions being observed for the latter. 
Such differences can, however, be expected, as the additional momentum built up during these rebounds is different: The bubble travels away from the free surface but cannot travel away from gravity.
Overall nonetheless, the second rebound energy, relative to the initial bubble energy, increases with increasing $\zeta$ in conjunction with the first rebound energy.

\begin{figure}
\begin{center}
\includegraphics[width=.48\textwidth, trim=0cm 0cm 0cm 0cm, clip]{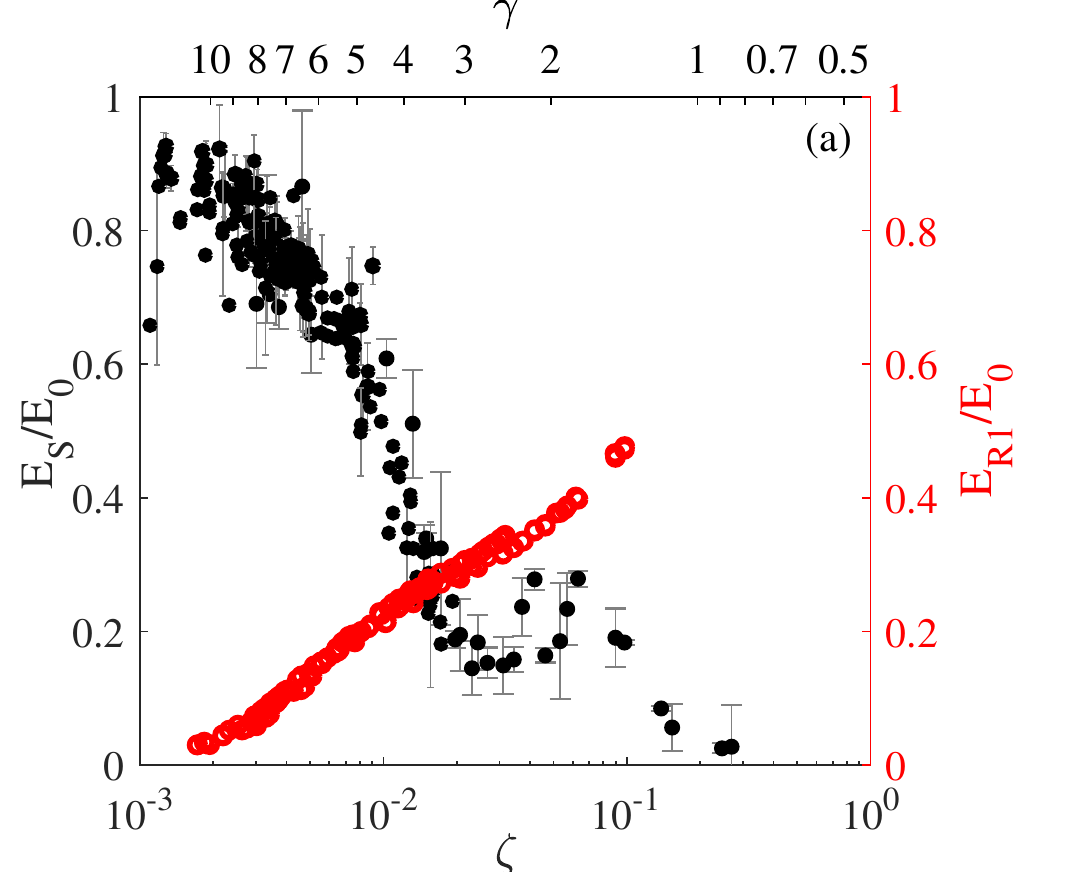}
\includegraphics[width=.48\textwidth, trim=0cm 0cm 0cm 0cm, clip]{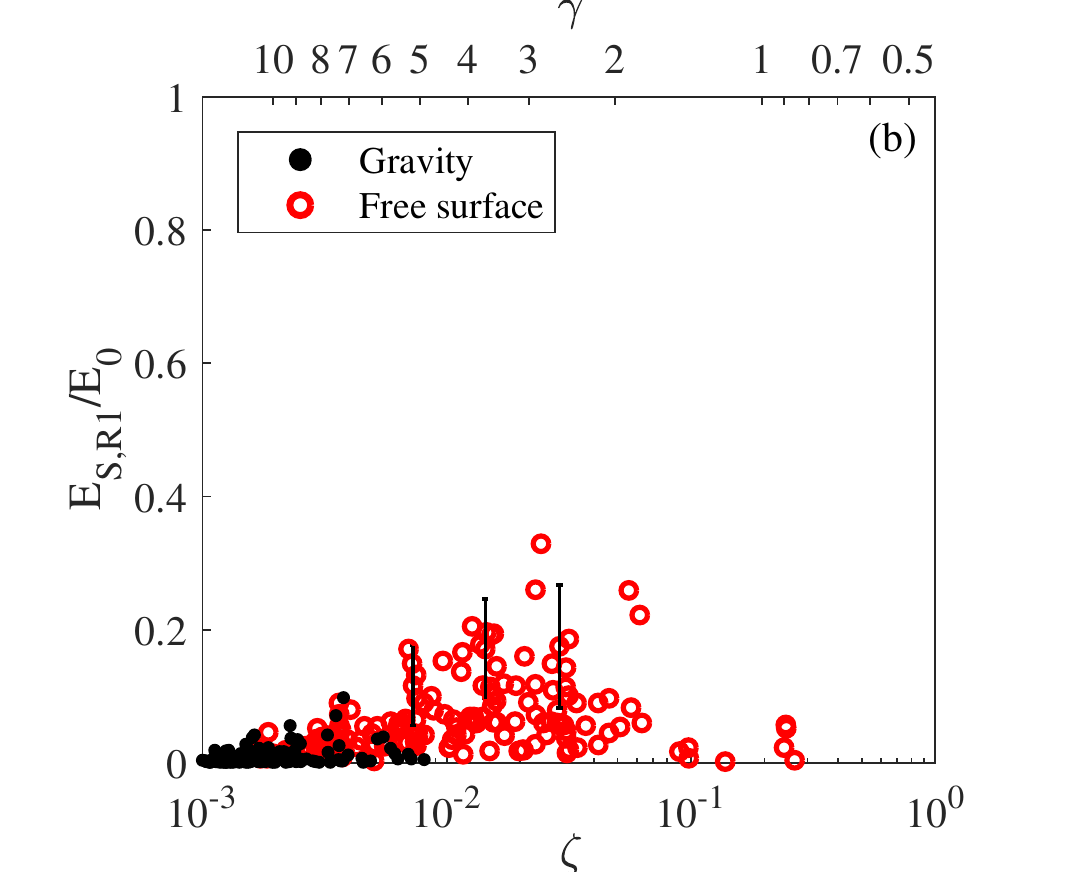}
\caption{(a)~Shock wave and rebound energy, normalized to initial bubble energy $E_{0}$ as a function of $\zeta$, for a bubble deformed by a neighboring free surface (shock energy data replotted from Ref.~\cite{Supponen2017a}). (b)~Rebound collapse shock wave energy, normalized to initial bubble energy $E_{0}$, as a function of $\zeta$ for bubbles deformed by gravity and by a neighboring free surface. The equivalent stand-off parameter $\gamma$ is shown on the top axes for comparison. Each data point corresponds to a single collapse event. Some of the measurement uncertainties are shown by the error bars.}
\label{fig:fig5}
\end{center}
\end{figure}

The energy of the shock wave emitted at the bubble collapse is computed by assuming a spherically symmetric shock and using the relation $E_{S} = 4\pi d^{2}(\rho c)^{-1}\int p(t)^{2}dt$~\cite{Vogel1988,Supponen2017a}, where $d$ is the radius of the spherical shock wave as it reaches the hydrophone, $c$ is the sound speed in the liquid, $p(t)$ is the calibrated pressure signal recorded by the hydrophone, and the integration bounds are selected to comprise the full waveform of the shock wave produced at collapse.
More details on the calibration of the voltage signal into pressure and thereby energy can be found in Ref.~\cite{Supponen2017a}.
Here, the bubbles are deformed by a near free surface to reach higher levels of deformations and thus bigger rebounds.
The energy of the shock waves emitted at the first bubble collapse is displayed in Fig.~\ref{fig:fig5}(a) as a function of $\zeta$ (replotted from Ref.~\cite{Supponen2017a}) and compared with the rebound energy of the same bubbles as computed from the hydrophone signal.
As less energy is radiated away by the shock waves with increasing $\zeta$, the rebound is able to carry more energy. 
Most of the remaining energy is dissipated away through the flow associated with the microjet formation, while energy dissipation through heating is expected to be negligible~\cite{Supponen2017b}.
Figure~\ref{fig:fig5}(b) displays the total energy of the shocks emitted at the collapse of the first rebound bubble, $E_{\rm S,R1}$, normalized to the initial bubble energy $E_{0}$.
The shock wave emission process from the collapse of the strongly deformed rebound is highly irregular, resulting in the strong scatter in the figure.
However, a maximum of the rebound shock energy may be observed at $\zeta \approx 0.02$--$0.03$. 
As a result of the rebound bubble gaining more energy with increasing $\zeta$, up to $\zeta \approx 0.03$, its own collapse is able to generate more energetic shock waves.
However, as the rebound bubble becomes more deformed with a further increasing $\zeta$, the shocks emitted at its own aspherical collapse lose energy likewise to the primary bubble.
It is apparent that the energy of the rebound shock(s) can reach values similar to - or even higher than - the first collapse shock(s) at $\zeta>0.02$; at $\zeta\sim0.03$, the energy of the latter is less than 0.2$E_{0}$, while here the second rebounds are able to emit shocks with energies exceeding 0.3$E_{0}$. 
The results agree rather well with previously presented findings for bubbles near rigid boundaries  by Vogel and Lauterborn~\cite{Vogel1988}.
They showed that for bubbles collapsing at a certain range of distances to a solid boundary ($0.5<\gamma<2$, or equivalently, $0.05<\zeta<0.8$), the shocks from the rebound collapse may carry a comparable amount of energy to the fist collapse. 
Their range is slightly different compared to our findings, which is explained by the rebounds from bubbles collapsing near rigid surfaces being migrated toward the boundary while here they are repelled by the free surface.


\section{Discussion}

The formation of the rebound bubble naturally must depend significantly on the gas inside the collapsing bubble, which we have not varied in a controlled way in this work.
Varying $\Delta p$ likely also varies this gas content, which varies the energy partition into the rebound for spherical collapses~\cite{Tinguely2012} and could be the reason for bubbles in lower $\Delta p$ consistently producing slightly bigger rebounds in Fig.~\ref{fig:fig4}.
Indeed, the gas inside the bubble, in our experiment, likely comprises 
(i)~laser-generated gas, as has been reported in the past~\cite{Sato2013} - we assume this gas pressure to be proportional to the energy deposited by the laser to generate the bubble; 
(ii)~noncondensible gas, for which we assume the partial pressure to be proportional to the bubble volume; 
(iii)~diffused gas, for which we assume the partial pressure to be proportional to the total exposed bubble surface being covered during its lifetime; and 
(iv)~vapor, the partial pressure of which is assumed to be the vapor pressure of the liquid, $p_{v}$, during most of the bubble's lifetime, but which may act as a noncondensible gas in the final collapse stages due to its condensation rate not being able to keep up with the bubble's rapid volume reduction~\cite{Fujikawa1979}.
Much more thorough modeling is needed to obtain proper predictions of the gas effects on the rebound dynamics.
However, as $\zeta>10^{-3}$ (equivalently, when $\gamma<14$), the effect of deformation starts to dominate the rebound dynamics over the effect of the different $\Delta p$, as demonstrated in Fig.~\ref{fig:fig4}.
Why exactly there is a logarithmic increase of the rebound bubble energy as a function of $\zeta$ is a question yet to be answered.
Perhaps finding out how the kinetic energy of the liquid contributing to the microjet formation behaves as a function of $\zeta$ could give some insight into this logarithmic trend.
Nevertheless, these results imply that care should be taken with theoretical models aiming to accurately model the rebound dynamics of bubbles that are not perfectly spherical, which is the case with most bubbles in reality.
Furthermore, as was shown in Fig.~\ref{fig:fig5}, there is a range of $\zeta$ in which the rebound bubbles are both large and spherical enough to emit relatively strong shocks upon their collapse.
In addition, if the rebound bubbles find themselves closer to a rigid boundary due to the bubble migration toward it, the damaging effects may be even stronger.
Therefore, the prediction of the rebound bubble's behavior is invaluable in determining the damaging effects cavitation bubbles can produce.
The rebound collapse somewhat depends on the origin of the initial deformation, as already indicated by the second rebound energies in Fig.~\ref{fig:fig4}.
It would be interesting, in the future, to define and evaluate a parameter describing the anisotropy felt at the second bubble collapse.

\section{Conclusion}

The present study experimentally quantifies the effect of bubble asymmetry on the energy of the rebound bubble.
We find a logarithmic increase of the first rebound energy, in normalized coordinates, as a function of the initial anisotropy parameter $\zeta$ in the range $10^{-3}<\zeta<10^{-1}$ regardless of the origin of deformation (gravity or neighboring free surface) and bubble radius, with a negligible dependence on the driving pressure.
The second rebound bubble takes between 10\% and 35\% of the first rebound, depending somewhat on the origin of deformation and with no clear variation with the initial $\zeta$.
Our findings may be valuable as a reference for analytical and numerical modeling of multiple oscillations of cavitation bubbles.

\acknowledgements{We gratefully acknowledge the support of the Swiss National Science Foundation (Grants No.\ 200020-144137 and P2ELP2-178206), the University of Western Australia Research Collaboration Award (PG12105206) obtained by D.O.\ and M.F., and the European Space Agency. We thank Philippe Kobel and Nicolas Dorsaz for their valuable help with the experiment.}

\bibliography{bibliography}

\end{document}